\documentclass[conference]{IEEEtran}

\ifCLASSINFOpdf
\else
\fi
\usepackage{pdfpages} %
\usepackage{array} %
\usepackage{multirow} 
\usepackage{dcolumn} %
\usepackage{url}

\usepackage{hyperref}
\usepackage{tabularx}
\usepackage{multirow}
\usepackage{amsmath}
\usepackage{amstext}
\usepackage{amssymb}
\usepackage{caption}
\usepackage{xspace}

\begin{document}
%
\title{Crime Scene Re-investigation: A Postmortem Analysis of Game Account Stealers' Behaviors} 

\author{
\IEEEauthorblockN{Hana Kim}
\IEEEauthorblockA{
Korea University\\
Seoul, Republic of Korea\\
hanada@korea.ac.kr}

\and
\IEEEauthorblockN{Seongil Yang}
\IEEEauthorblockA{
ETRI\\
Daejeon, Republic of Korea\\
siyang@etri.re.kr}

\and
\IEEEauthorblockN{Huy Kang Kim}
\IEEEauthorblockA{
Korea University\\
Seoul, Republic of Korea\\
cenda@korea.ac.kr}}

\maketitle

\begin{abstract}
As item trading becomes more popular, users can change their game items or money into real money more easily.
At the same time, hackers turn their eyes on stealing other users’ game items or money because it is much easier to earn money than traditional gold-farming by running game bots.
Game companies provide various security measures to block account-theft attempts, but many security measures on the user-side are disregarded by users because of lack of usability. In this study, we propose a server-side account theft detection system
base on action sequence analysis to protect game users from malicious hackers. We tested this system in the real Massively Multiplayer Online Role Playing Game (MMORPG). By analyzing users’ full game play log, our system can find the particular action sequences of hackers with high accuracy. Also, we can trace where the victim account’s stolen money goes.

\end{abstract}

\begin{IEEEkeywords}
Account theft, User behavior analysis, Sequence analysis, MMORPG
\end{IEEEkeywords}

\IEEEpeerreviewmaketitle

\section{Introduction}
As the Internet has increasingly become a big part of people's daily life. Also, real economy and virtual economy combines fast then the border between real and virtual world becomes blur. 
Hackers develop various attack tactics to take advantage of the online world; Account theft is one of the frequent attacks. If a hacker's can steal a user's authentication information (e.g. ID and password), then he can gain personal information and cyber assets possessed by the user. From this point of view, hackers turn their eyes on online game companies and game money exchange sites (e.g. ItemBay\footnote{http://www.itembay.com} and ItemMania\footnote{http://www.itemmania.com, ItemBay and ItemMania raise a profit of 40 million dollars out of commission fees for the trades annually.}), because those sites' security level is relatively lower than government or banking sites. Also, their platform make it easy for users to turn their items into cash. 
For example, in case of Blizzard, a game producer of Overwatch and World of Warcraft, was hacked in 2012; Unfortunately, Blizzard users' information was leaked \cite{IEEEhowto:news2}. In case of Steam, a famous game delivery platform, reported that almost 77,000 users' information is illegally leaked every month \cite{IEEEhowto:news3}. 

\begin{figure}[!h] 
\centering
\includegraphics [width=2.5in]{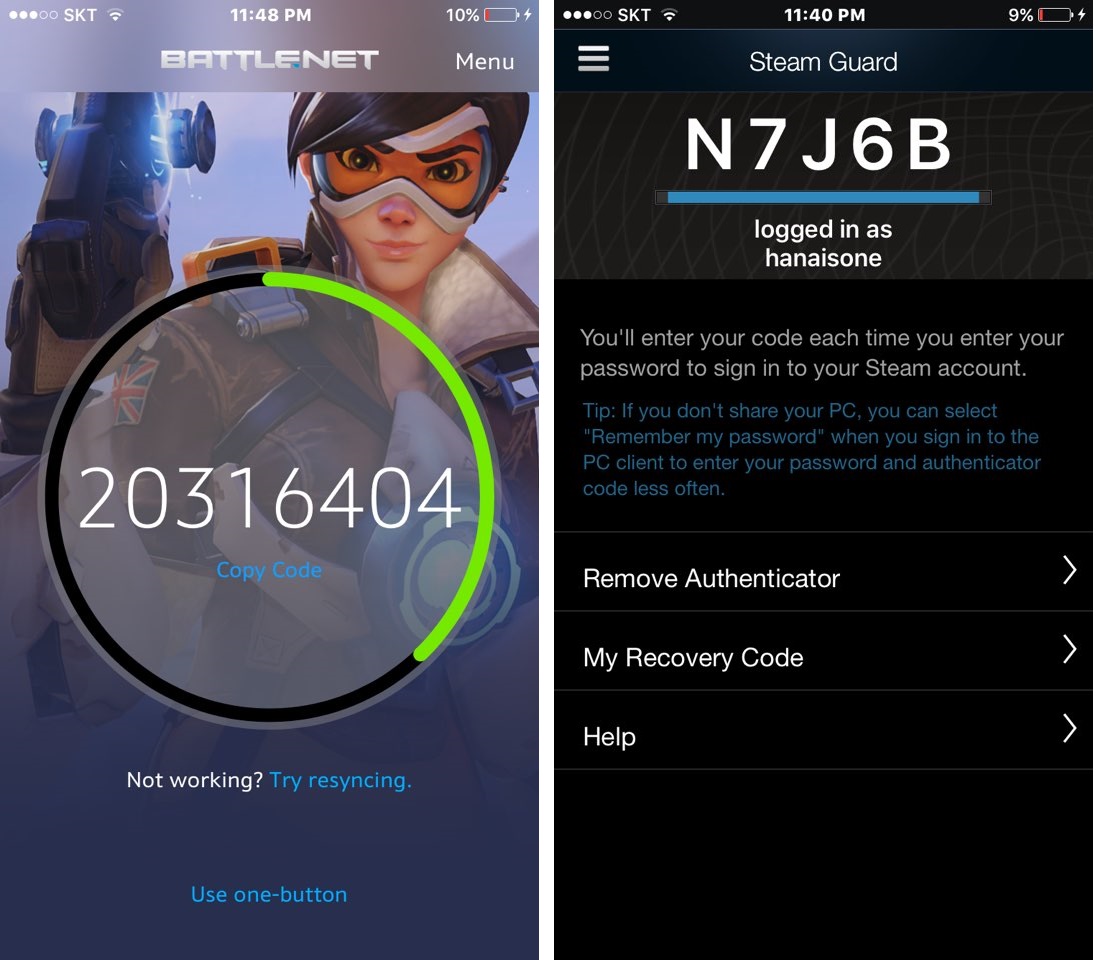} 
\caption{Blizzard and Valve's mobile authentication app.}
\label{fig:fig_0}
\end{figure}

To respond this kind of attack, many game companies are providing various security measure to protect users from account theft attacks. 
Typical examples are OTP (One Time Password) for strong authentication, machine ID inspection for detecting a login from unusual machines, Antivirus programs to detect keylogger or password stealer program. 
Even though many game providers offer a variety of services to prevent account theft for free, but most of them are not widely used because of its inconvenience. For example, many game companies provide OTP authentication software for PC or mobile, but users disregard this security protection measure, because they do not want to install additional authenticator program on their devices. Figure \ref{fig:fig_0} shows the screenshot of the mobile OTP apps provided by Blizzard and Valve. 

In this paper, we analyzed the action sequences of the account thieves and proposed a model to detect account thieves based on the analysis results. The proposed detection model is useful in detecting the theft of users even if the users do not perform security measures at the user-side. We analyzed transaction networks of the account thieves and analyzed their transaction characteristics and analyzed whether they are related to game bots.


\subsection{Contributions}
To demonstrate the feasibility of the proposed account theft detection method, we use the ground truth data from one of the largest MMORPGs, Aion\footnote{http://na.aiononline.com/en/}, developed by NCSoft. The key contributions of this study are as follows: 
\begin{itemize}
\item\texttt{} We conducted real MMORPG dataset analysis to detect hacker's behaviors. To this end, we did action sequence analysis method to detect an account theft attempt in MMORPG. Our proposed method can detect account theft with high accuracy.  
\item\texttt{} We reveal account thefts are highly correlated with game bots and GFG; The victimized accounts' money transfered to GFG's accounts.
\end{itemize}

\subsection{Ethical Considerations}
We note that all collected log data for this study satisfy related ethical review. Before installation, all game users were asked to acknowledge a consent form under the End User License Agreement (EULA) and domestic laws, which informed them that their data might be used in improving the quality of the installed game. Anonymous data samples were collected confidentially only for the purposes of conducting statistical analyses. 

\subsection{Organization}
The remainder of this paper is as follows. In \textsection\ref{section:background}, we introduce terminologies and background knowledge used in this paper. In \textsection\ref{section:related}, we review  related works. In \textsection\ref{section:method}, we propose a method to detect account theft based on action sequence analysis. We demonstrate the overall detection process and evaluate the performance of the proposed method. Also, we reveal the trace result by analyzing item trade network. Finally, we summarize our findings and conclude in \textsection\ref{section:conclusion}. 


\section{Background}
\label{section:background}

\subsection{Terminology}
\label{section:term}
\begin{itemize}
\item\texttt{}{\bfseries Account theft, Account thief and victim account.} Account theft is a criminal activity to steal a user's account (ID and password) for abusable purposes. Account thief and victim account is an attacker and victim in the course of account theft, respectively. 

\item\texttt{}{\bfseries Game Bot.} Game bot is an AI program designed to play a game automatically without human's control. 
This program discourages normal users of the program to play a game since they feel deprived of their opportunities to win to their counterparts using the program. 

\item\texttt{}{\bfseries GFG (Gold Farming Group).} GFG is an industrialized group which run lots of game bot to earn cyber money or item efficiently. Usually, GFG run game bots composed of three sub-groups: gold-farmer group to collect game items, merchant group to turn them into cash and banker group to keep the items-turned game money.

\item\texttt{}{\bfseries RMT (Real Money Trading).} RMT is an activity of trading cyber item or money for real money. By doing RMT, users can gain high-value game asset by paying real money. Therefore, it is possible to reach the highest level fast with less efforts in game. 

\item\texttt{}{\bfseries Game log.} Game server generates various types of log at the server-side in order to record users' in-game activities. In general, this log can be used for bug trace, user behavior analysis, and game bot detection. Usually, game log includes account ID, event or action ID, time stamp, and context-related information. 
\end{itemize} 

Traditionally, game bots run by GFGs and RMT are the most serious problems in online game. Because it can destroy the fair-play rules of game. Even worse, they can eventually lead the collapse of economic balance in game. Therefore, most game companies make efforts to detect and block game bots and RMT \cite{lee2016you}. 

\section{Related Work}
\label{section:related}
Many researches conducted on account theft detection methods by gathering additional information at the client-side (e.g. MAC address for machine ID detection, Antivirus program for detecting keylogger programs), or data mining techniques at the server-side. To provide better usability to users, client-side account theft detection is not well used today. In case of data mining approach at the server-side, most frequently method used features are game user's login time, IP address, MAC address and movement pattern. Chen and Hong \cite{chen2007user} proposes a model to detect the theft by using Kullback-Leibler Divergence (KLD) on a pattern of the playing time in an MMORPG. Oh \texttt{}{\itshape et al.} \cite{oh2012automatic} applies a statistical technique to data, like experience level, trade and playing time to come up with a model to detect. 
Woo \texttt{}{\itshape et al.} \cite{woo2012automatic} classified account theft process into exploration, monetization and theft. They used decision tree for classification of account theft. Choi \texttt{}{\itshape et al.}\cite{choi2011detecting} classified account theft type into three: ``quick in-and-out", ``cautious" and ``bold". Based on their three major account theft scenario, they verified the feasibility by using the neural network. They used game related features such as experience level, login time, and play duration. However, this study has several limitations as follows their proposed method showed high accuracy. However, once account theft pattern has changed, they need to train a neural network again. 
Lee \texttt{}{\itshape et al.} \cite{lee2015game} used action sequence analysis extracted from the game action logs in order to detect game BOT users.  


Most of the studies on account theft mainly focused on developing detection methods, but they used simulation data not real dataset. Also, their training dataset used small number of samples collected in a short term period. 
In our study, we conduct the action sequence analysis to create a detection model with real data collected in a long term period. Also, with our best knowledge, our study is the only study which traces the full money trail of the victim accounts. As a result, we can unveil where the stolen money goes and who are the wire-puller behind the scene. 


\section{Methodology}
\label{section:method}
\begin{figure}[!h]
\centering
\includegraphics [width=3.3in]{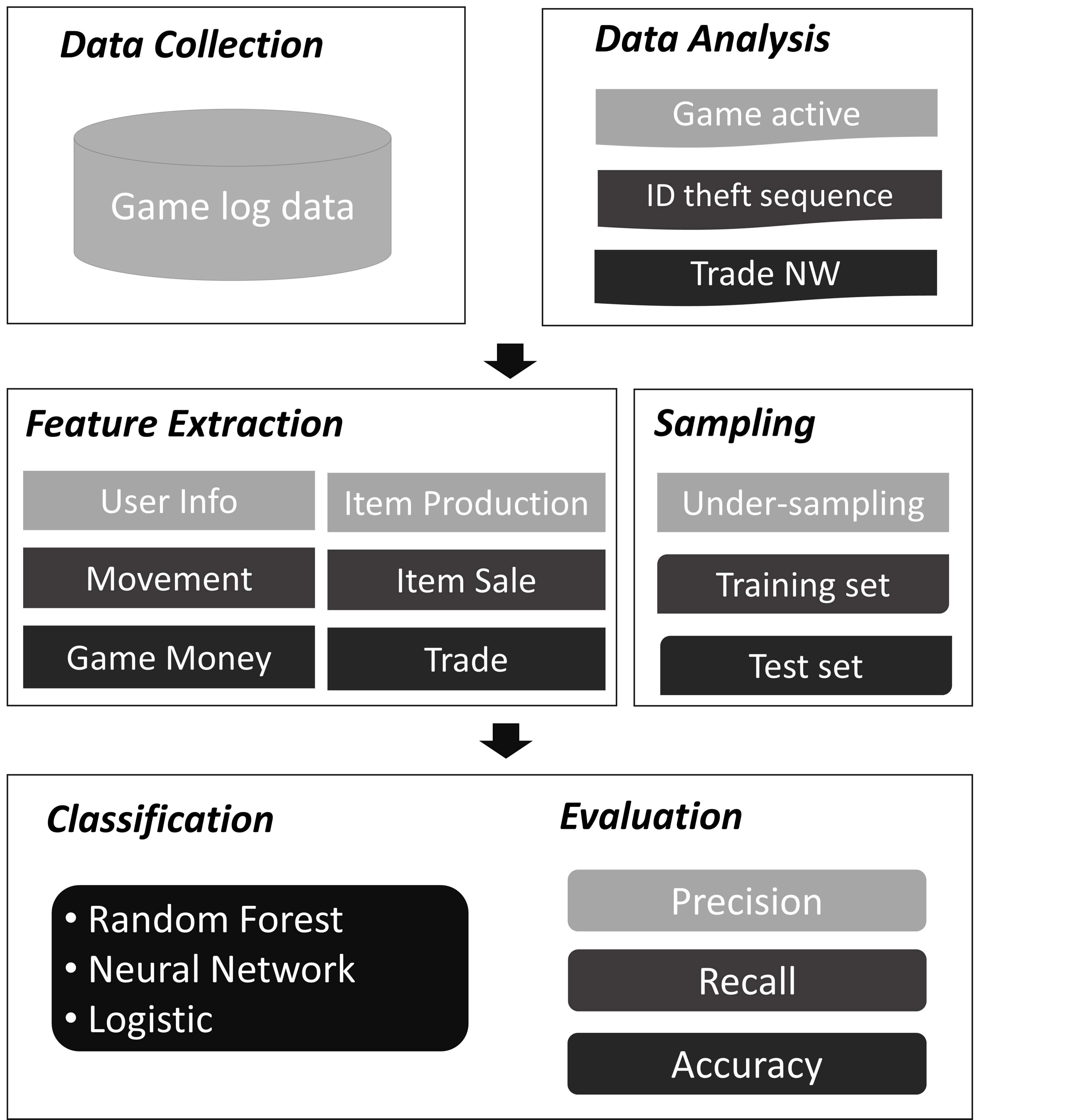}
\caption{Overall process of the proposed methodology}
\label{fig:fig_method}
\end{figure}

In this section, we propose a method to detect account theft. Also, we did postmortem analysis on the victim accounts. 
Figure \ref{fig:fig_method} shows the overall process of the proposed method. 

\subsection{Dataset}
We used full logs collected from one of the Aion servers from June 25th to July 4th in 2010. In total, there were 23 confirmed account theft cases (i.e. ground-truth data). During this period, 1.1 cases occurred on average; and the peak number of case was 11 cases in the same day. Most of account theft cases occurred in a row. 
We built the blacklist by extracting IP addresses used by victim accounts and then we extracted all game logs from the server related to the users who were connected from the blacklist IP addresses at the same date when the accounts theft occurred. In total, there were 82 times of login and logout events from the 9 suspicious IP addresses. 

To reduce the size of logs, we excluded 18 types of logs which were automatically generated by game system. The excluded logs include less informative data such as game maintenance notice. Finally, we carefully chose 41 types of logs for the analysis such as combat, trade, movement and hunting. Absolutely, the login and logout related logs were fully included to avoid information loss. 

\subsubsection{Ground truth}
In many previous studies, some researchers used data sets generated artificially by hired players and hacking programs. Other researches used a list of  accounts that other players reported them as suspicious players. It may generate false positive errors. 
The ground truth data (detected bot users, GFGs, and account theft cases) used in this paper is confirmed list by NCSoft. These ground truth data was carefully collected by professional monitoring persons called as GMs (Game Masters) who can secretly monitor the game world and detect malicious users. 
Also, all detected malicious users can raise a petition to appeal they are innocent but the detected users did not claim. 

\subsection{Analysis of Account Thieves' Behavior}
\begin{figure}[!h]
\includegraphics [width=3.4in]{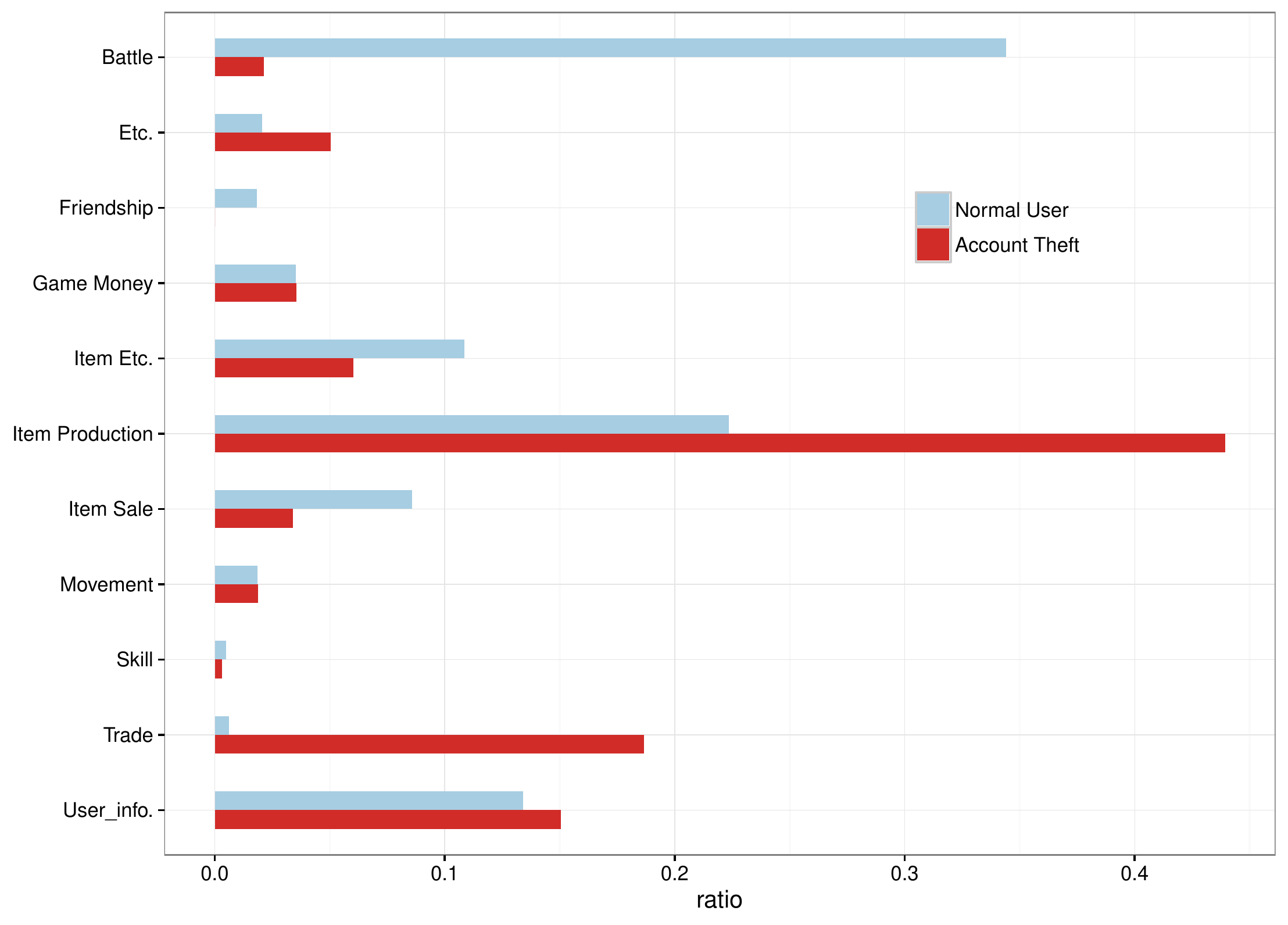}
\caption{Action Ratio by Normal Users and Account thieves}
\label{fig:fig_1}
\end{figure}

In order to understand the behaviors of account thieves, 11 different types of actions are reviewed: battle, skill, friendship, trade, item production, item sale, item etc. (miscellaneous item related events), game money, movement, user information and etc. Figure \ref{fig:fig_1} shows the comparison of action ratio between normal users and victim accounts (played by account thieves) for each type of action. Victim  accounts have smaller number of friendship or battle related actions;but, they have higher number of trade and item related activities than normal users. 

We also analyzed the expenditure ratio of the victim accounts in a week-long basis. The expenditure ratio is defined as a ratio of spent money and possessed money per day. 
The expenditure ratio is calculated as shown in Equation (1) below.\vspace{+0.1cm}

\scriptsize
\begin{equation}
\centering
\frac{Daily \,used \,money}{Amount of money \,at \,the \,beginning\,+\, Daily \,used \,money}
\end{equation}\vspace{0.1cm} 

\begin{figure}[!h] 
\centering
\includegraphics [width=3.3in]{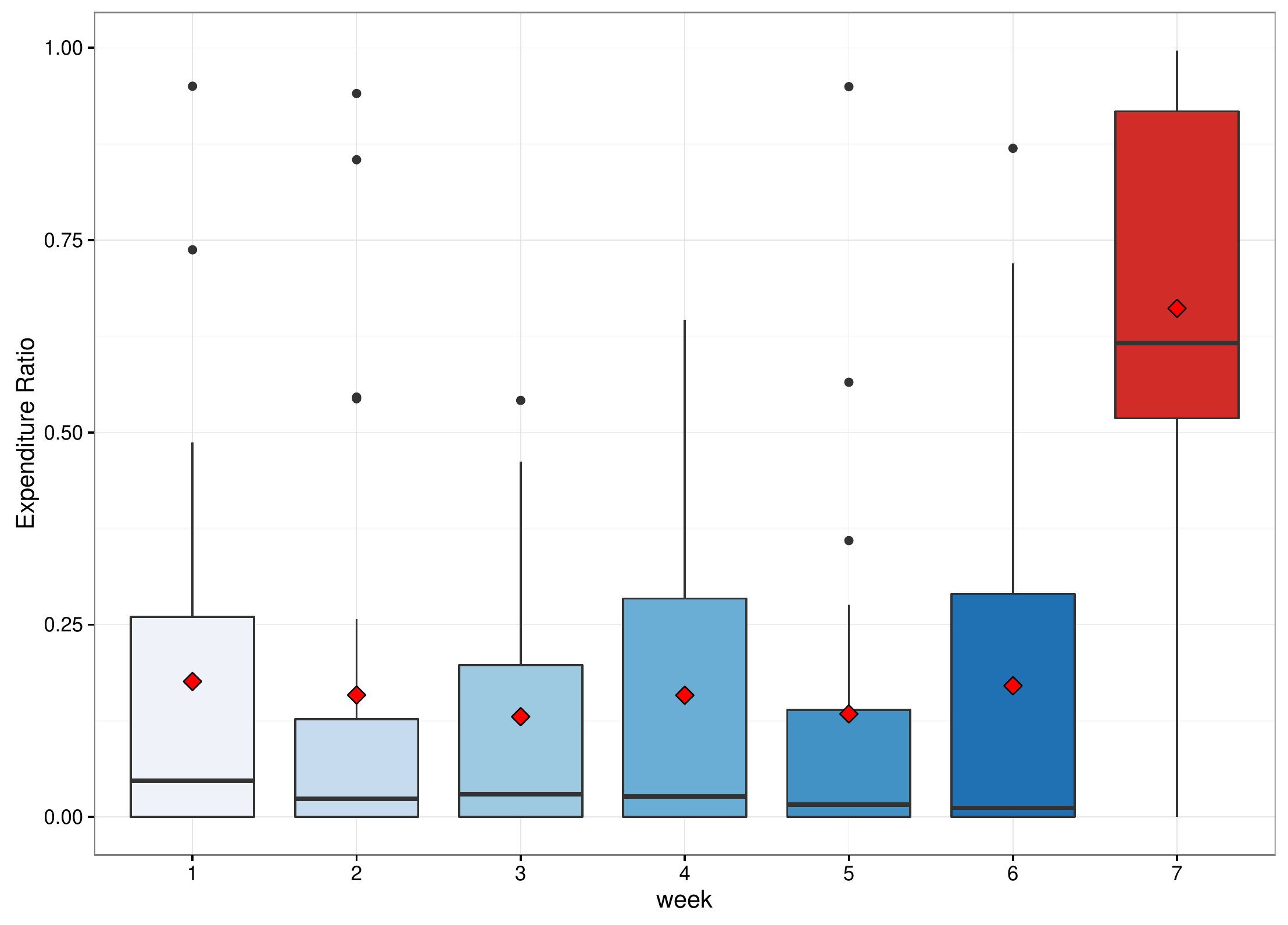} 
\caption{The weekly expenditure ratio of the victims (account theft occurred in the day 7 in X-axis.)}
\label{fig:fig_2}
\end{figure}

\normalsize
Figure \ref{fig:fig_2} indicates that most of the expenditure ratio shows the ranges around 20\% to 25\% during the first-six days when no account theft occurred. It goes up to 70\% on the day when account theft occurred. 
In this box-plot, Red `diamond' means an average value, `straight line' means a median value and `box' means an expenditure ratio. We can find the account thieves main purpose is to steal game money or items from victim accounts; then, they sell the stolen items in order to gain real money. Therefore, the victim accounts played by account thieves show higher expenditure ratio than normal users. 

Through the comparative analysis of action types and expenditure ratio, we can find the account thieves main purpose. 
Based on this findings, we perform action sequence analysis to find common action sequences (that can be used as a detection signature) and visualize account thieves' action pattern more clearly.  

\subsection{Action Sequence Analysis of Account Thieves'}
Table \ref{table:table1} shows the mapping for the action sequence analysis. 
First, we categorize all action behaviors into 11 categories. Then, we mapped 11 different action categories with an alphabet character from `A' to `K'. Each category includes the related actions as follows: For example, `User information' includes the actions related to user's level, playtime, and experience point. `Movement' includes all movement related actions such as teleport, movement, or flight. `Item Purchase' and `Item Sale' include the events when an item is purchased or sold through an NPC shop or a user's own shop. `Trade' includes all events when a user exchanges or transfers items or money.

\begin{table}
\centering
\caption{Action categories and their mapped alphabet characters for the sequence analysis}
\label{table:table1} 

\begin{tabular}{|>{\centering}m{2cm}|c|>{\centering}m{2cm}|c| } \hline
 {\bfseries Category} & {\bfseries character} & {\bfseries Category }& {\bfseries character}\\ \hline 

 Login                              & A  & Item Production & G \\   \hline 
 Logout                            & B  & Item Purchase  & H \\   \hline 
 User Information               & C & Item Sale         & I \\   \hline
 Movement                       & D & Trade  & J \\   \hline 
 Decrease of Game Money & E & Etc. & K \\   \hline 
 Increase of Game Money & F & -                     & -\\   \hline 
\end{tabular} 
\end{table}

\begin{table}
\centering
\caption{A ratio of the account thieves' playing times}
\label{table:table2} 

\begin{tabular}{| >{\centering}p{0.2cm} | >{\centering}p{1.5cm} | c | c | m{2.5cm} | } \hline
 \multicolumn{2}{ |c| }{{\bfseries Type}} & {\bfseries Count} & {\bfseries Ratio} & {\bfseries Note} \\ \hline 
 1 & Less than 1 minutes   & 32  & 39\% &     \\   \hline 
 2 & Less than 5 minutes   & 33  & 40\% &     \\   \hline
 3 & Less than 10 minutes & 7    & 7\%  &     \\   \hline 
 4 & More than 10 minutes & 10  & 10\% & Within an hour: 8 More than an hour: 2 \\   \hline 
\end{tabular}
\end{table}

\begin{figure*}[!h]
\centering
\includegraphics [width=6in, height=3.8in]{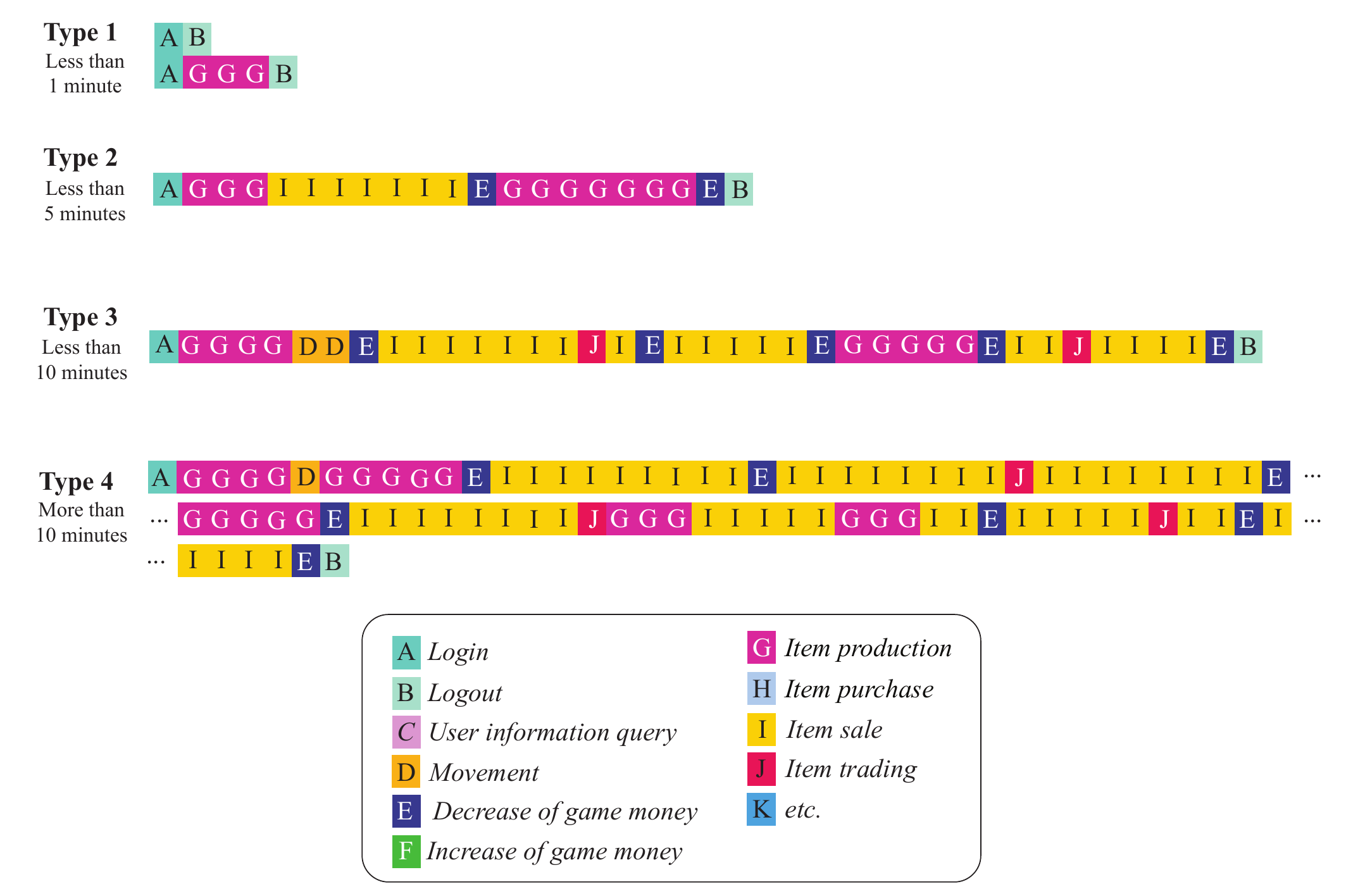}
\caption{Common action sequence of victim accounts}
\label{fig:fig_3}
\end{figure*}

For the detected 82 times of login and logout events, we analyzed the playing time distribution for each login case. Table \ref{table:table2} shows a ratio of game-playing time for each login session. We categorized the game-playing time into 4 types. It indicates that 88\% of victim accounts (Type 1, 2 and 3) are abused within 10 minutes. That means account thieves play the victim account within 10 minutes after login. (Thus, they can steal items and sell them within 10 minutes.)
In order to understand what actions takes place for each type, the account thief's action sequences are visualized in a temporal view. 
Figure \ref{fig:fig_3} shows an action sequence graph of victim accounts by type. 
We extract and display the common action sequences commonly existed in all user's action sequences for each type. 

\begin{itemize}
\item\texttt{}\textbf{Type 1} `Type 1' victims in Table \ref{table:table2} show that Login \textit{(A)} and Logout \textit{(B)} actions are repeated. It seems automated and repeated procedures by the account thieves to check whether the account's original user can be aware of their illegal login actions. Because many modern online games provide notification when suspicious login events occurred. When account thieves confirm that there are no responsive checking events by account's original user, then they proceed to steal victim's possessions. We can discover the action sequences \textit{(G)-(G)-(G)-(B)} are following. It means account thieves utilized the victim's money to produce items (i.e., pink colored (G)) and then just did logout to check whether they are detected or not.  

\item\texttt{}\textbf{Type 2} `Type 2' victims shows that items are sold by account thieves; We can find `Type 2' victims commonly have the action sequences of \textit{(G)-(G)-(G)-(I)-(I)-(I)-(E)}. That means account thieves produce items then sell them and use the earned money. Yellow colored \textit{(I)} means `Item sale', blue colored \textit{(E)} means `Decrease of Game Money' as defined in Table \ref{table:table1}.
\item\texttt{}\textbf{Type 3 and 4} `Type 3 and 4' victims basically repeat the common action sequences of Type 2 victim. Then, they additionally do `Trade'. (Red colored \textit{(J)} means `Trade'.) To summarize, once account thieves take control of the victim account securely, they repeatedly produce item by using victim's possessions and sell or trade the produced items to gain money. That's the reason why Type 3 and 4 victims repeatedly have sequences composed of I and E. 
\end{itemize} 

\subsection{Experiments and Evaluations}
On the result of the sequence analysis, some features are selected with unnecessary logs taken off. The selections are shown in the Table \ref{table:table3}. 95\% of all victim accounts includes higher than 40 in level (the highest level was 50 when the dataset was built). 88\% of users starts playing a game within 10 minutes. All accounts have zero in experience level beside one. (supposed that they are all used for profiteering activities, not for anything else. So, they might have 0 in experience point with the one account being used for simple item collection). Logs to movements are included since stolen items, or game money would be passed on to somewhere. A game money decreases immediately after it increases as part of the analyzed sequence. So, logs to the increase or decrease are included as well. Logs related to item production are one of the considerations to be selected since it is an important step in the account thieves' activities for undue profits, and item sale and trade included.

\begin{table}
\centering
\caption{Selected features for detecting account thieves}
\label{table:table3} 

\begin{tabular}{| c |m{5.5cm} |}
 \hline
 {\bfseries Category} & {\bfseries Feature}  \\   \hline 
 User Information & Level, Playtime, Experience point\\   \hline 
 Movement         & Number of movements \\   \hline
 Game Money    & Decrease the number of game money, Decrement of game money, Increase the number of game money, Increment of game money \\   \hline
 Item Production & Gain, Collection, Production, Item installation, Item non-installation, Extraction, Number of extract \\   \hline 
 Item Sale         & NPC shop, User shop, sales agency  \\   \hline 
 Item Trade       & Trade, move an item from a location to another  \\   \hline 
\end{tabular} 
\end{table}

As the number of account theft cases was noticeably smaller than normal users, this study constructed learning data after executing under-sampling technique of normal data and trained them using 10-folds cross-validation method. Three data mining methods are used as well: Multilayer Perceptron (MLP), Logistic and Random Forest. MLP is a method to construct weights on the node networked with a neural algorithm and changeable links and use repetitive learning to create a model. As one of regression analysis, Logistic can help better interpret and predict a model as it selects necessary variables. Random Forest learns multiple decision tree models in the process of training and then comprehensively looks at various results to sort them out.

\scriptsize
\begin{equation}
\centering
Precision = \frac{TP}{TP+FP} \,\,\, \,\,\,  Recall=\frac{TP}{TP+FN}  \nonumber
\end{equation}

\begin{equation}
\centering
Accuracy = \frac{TP+TN}{TP+TN+FP+FN}
\end{equation}
(where, TP: True Positive, TN: True Negative, FP: False Positive, and FN: False Negative) \\
\normalsize

This paper used three kinds of evaluation index (Precision, Recall, and Accuracy), and the definition of the three kinds of the index is shown in Equation (2). Precision means the ratio to whether detected results are actually account theft, and Recall is the ratio showing how much actual account theft is detected through the detection model. Accuracy is the index about how accurately the proposed model to all users predicted. 

Table \ref{table:table4} shows the results after those three methods used, with Random Forest performing the best at 84\%. 
Table \ref{table:table5} shows results from Random Forest on each type. Except for Type 2, all indicates higher than 0.9. In the case of Type 1 with less than 1 minute, once the type is detected, any attempts to take game money or an item can be preventable since the type can be interpreted as a preceding act for the actual purpose. 

\begin{table}
\centering
\caption{Performance of Account Theft Detection System (dataset in 2010)}
\label{table:table4} 

\begin{tabular}{| c |>{\centering} m{1.8cm} |>{\centering}m{1.8cm} |  c |}\hline
   &{\bfseries MLP} & {\bfseries Logistic} & {\bfseries Random   Forest}  \\   \hline 
 Precision & 0.676 & 0.710 & 0.844 \\   \hline 
 Recall & 0.665 & 0.682 & 0.835 \\   \hline
 Accuracy & 66\% & 68\% & 84\% \\   \hline
\end{tabular} 
\end{table}

\begin{table}
\centering
\caption{Performance of Random Forest For Each Type }
\label{table:table5} 

\begin{tabular}{| >{\centering} m{1cm} |>{\centering} m{1.5cm} |>{\centering} m{1.5cm}|c|}
 \hline
 {\bfseries  Type} &{\bfseries Precision} & {\bfseries Recall} & {\bfseries Accuracy}  \\   \hline 
 Type1 & 0.914 & 0.914 & 97\% \\   \hline 
 Type2 & 0.706 & 0.75 & 75\% \\   \hline
 Type3 & 1 & 0.714 & 71\% \\   \hline
 Type4 & 1 & 0.545 & 63\% \\   \hline
\end{tabular} 
\end{table}

The study includes an extra experiment with new Aion dataset from August 17th to 19th in 2015. The total number of login and logout is 42. The result follows as in Table \ref{table:table6}.

\begin{table}
\centering
\caption{Performance of Account Theft Detection System (new Aion dataset in 2015) }
\label{table:table6} 

\begin{tabular}{| c |>{\centering} m{2cm} |>{\centering}m{2cm} |  c |}\hline
   &{\bfseries MLP} & {\bfseries Logistic} & {\bfseries Random   Forest}  \\   \hline 
 Precision & 0.823 & 0.822 & 0.861 \\   \hline 
 Recall & 0.838 & 0.81 & 0.881 \\   \hline
 Accuracy & 84\% & 81\% & 88\% \\   \hline
\end{tabular} 
\end{table}

As a result, it shows better in accuracy than that of the dataset in 2015, with all higher than 80\%. Random Forest is set at 88\%, 4\% higher than that of the dataset in 2010. The findings help construe that the account theft has changed their patterns to the one as found in the sequence suggested in this study than before. 

The proposed detection model can discover victim accounts successfully. 


\subsection{Analysis of Account Thieves' Trade Networks}
In general, the `trade' in an online game means an activity to either exchange an item between users or pay or receive game money for the exchange. But, the network for trade by the account thieves is found to include trades where an item is only given without anything paid for it. Unlike it, 19\% of trades among general users includes the ones with money paid for exchange with the rest 81\% not paying for the exchange. But, as said, account thieves do not get paid for exchange 100\%. 

\begin{figure}[!h]
\centering
\includegraphics [width=3.3in]
{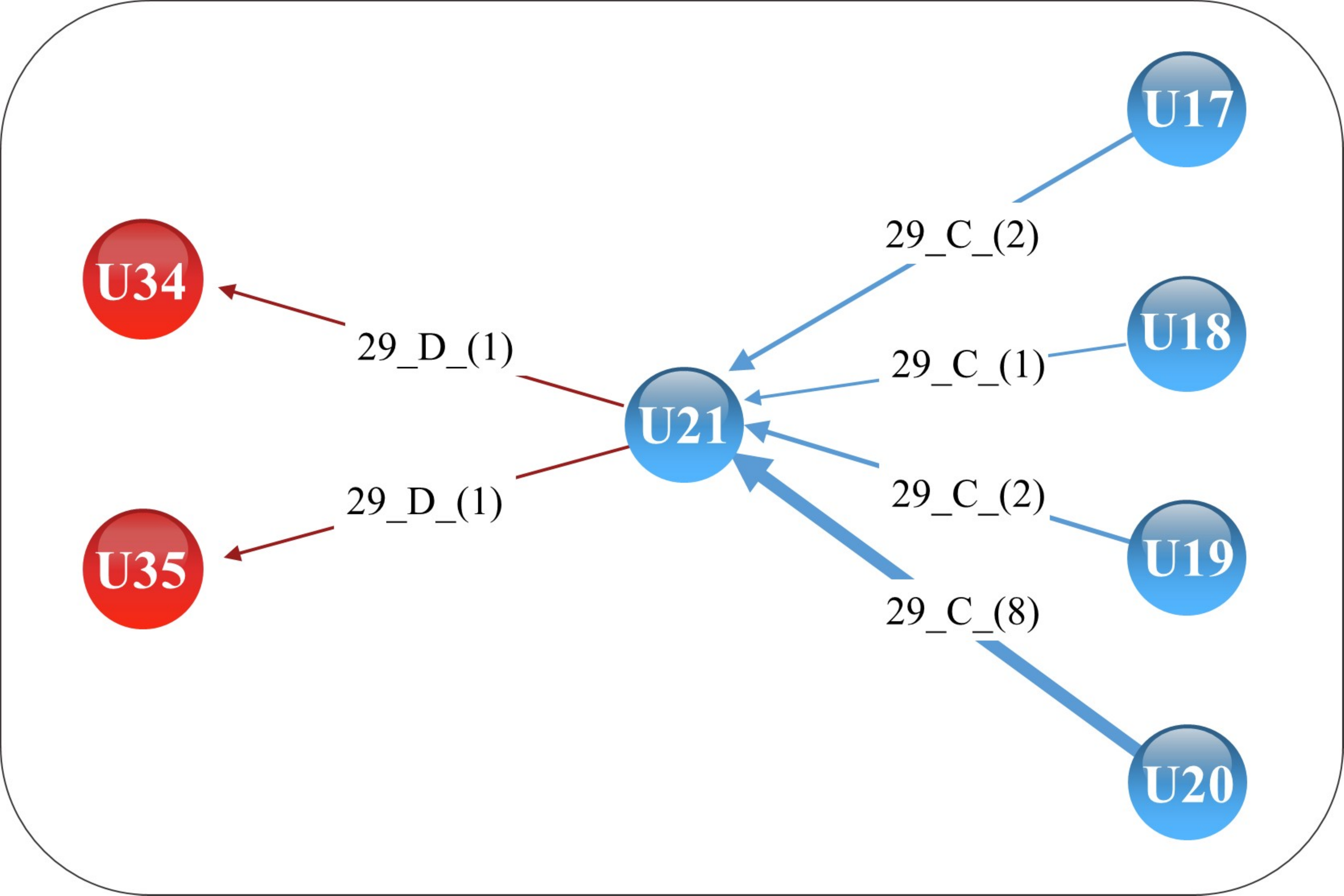} 
\caption{Patterns for trades by the abusers}
\label{fig:fig_4} 
\end{figure}

Two group are set in this study: first is ``account theft group" trading through IP addresses on a blacklist and second is ``suspicious group" trading not through them. The total number of trades with using victim accounts is 55, with 43 by the first group and 12 by the second group. Common trade patterns on the abused accounts are shown in Figure \ref{fig:fig_4}. The numbers in circles mean the id of users playing a game while the other numbers on each line indicate dates, locations of trade (Zone information in the game map), and number of trade. 

The victim accounts of U17 to 20 are used to send items or game money to another victim accounts of U21 at C in 29th. Then, U21 comes to D to give them to U34 and U35. This pattern is similar to 3 depth of GFG. This is because U21 stores all those stolen items or game money to a certain account and plays a role as a merchant tossing them to another account. U17 to 20 work as a gold-farmer collecting them. Lastly, U34 and U35 are a bank group. In order to see if there is any correlation with game bots, the study also traces any usage of the victim accounts and game bots. A trace of having used a game bot is found on all those accounts, indicating that it would be possible for the program to work to steel user information.   

\begin{figure}
\centering
\includegraphics [width=3.3in]{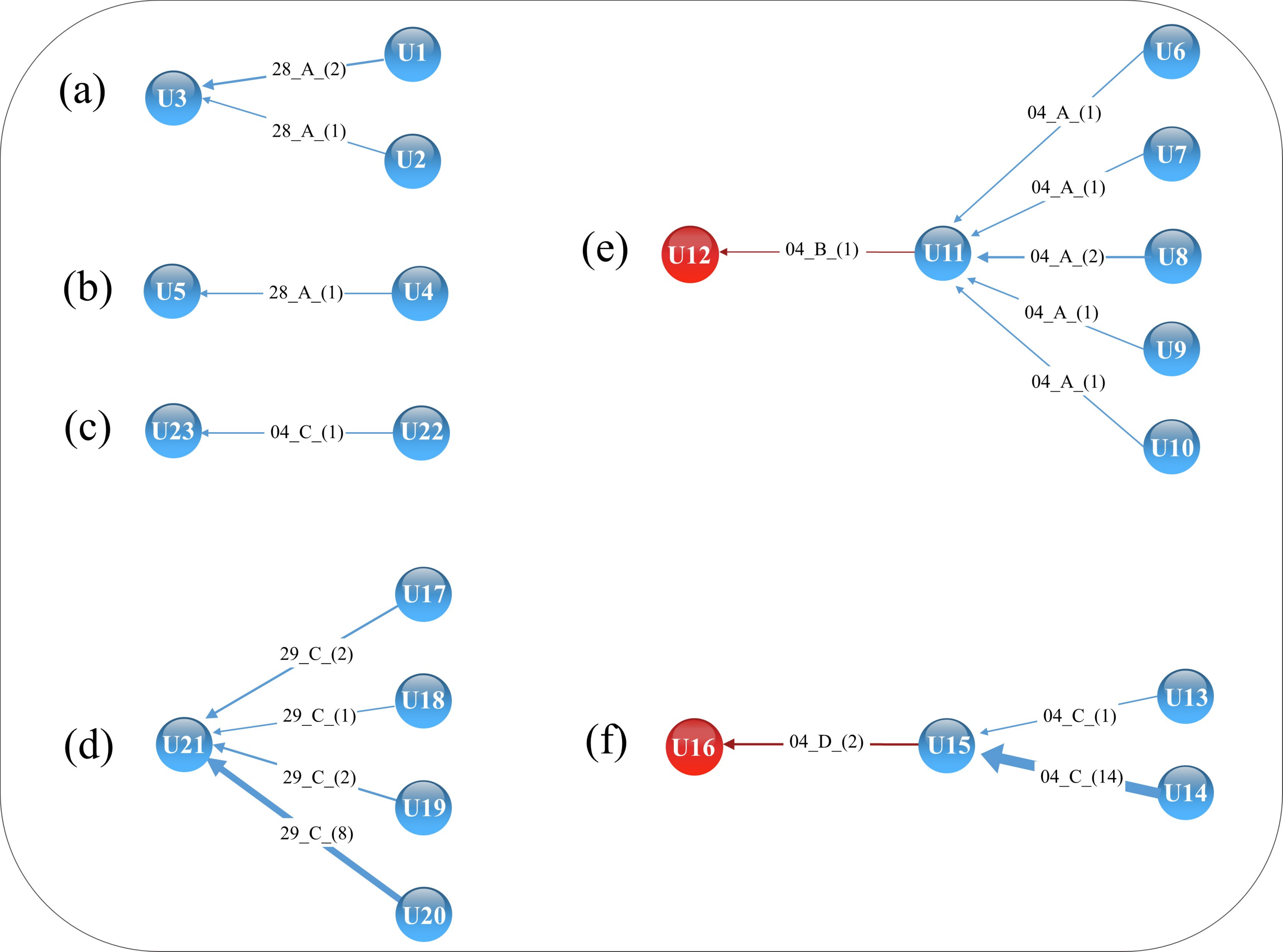} 
\caption{Trade network of the account theft group}
\label{fig:fig_5} 
\end{figure}

Figure \ref{fig:fig_5} shows the trade network of the account theft group. As explained above, specific characteristics of the trade network on the abused accounts can be found in `(d)', `(e)' and `(f)'. In this diagram, users on the right give items or game money to their counterparts on the left for nothing. The trade pattern goes on as they move from a place to another to send them to their final destination. Locations used for this kind of trades are fairly limited. That means, they prefer secluded places for the trades. `A', `B', `C' and `D' are used, with `A' and `B' being used to receive them coming from victim accounts and `B' and `D' being used to transfer them.

\begin{figure}[!h]
\centering
\includegraphics [width=3.3in]
{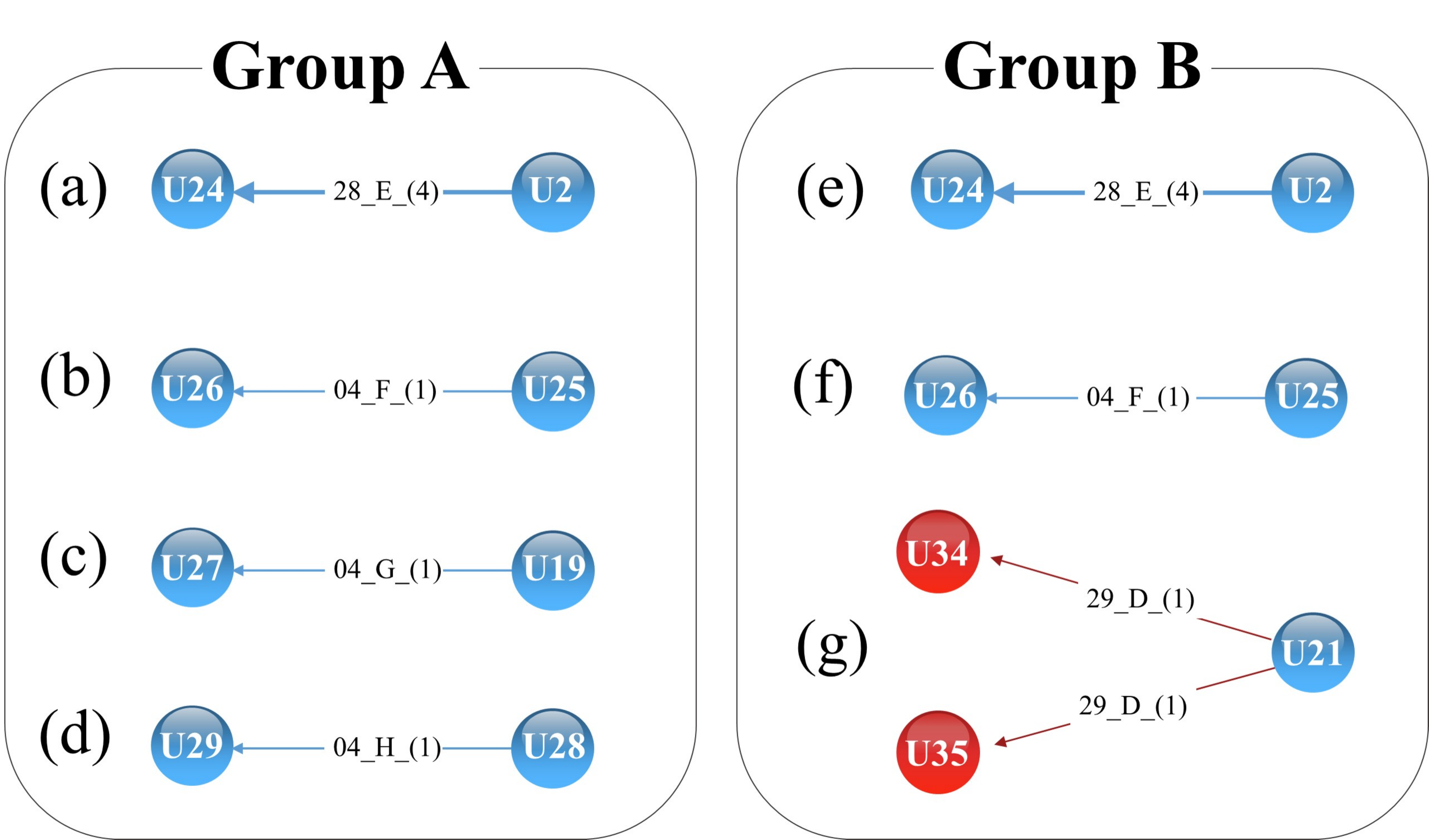} 
\caption{Trade network of the suspicious group}
\label{fig:fig_6} 
\end{figure}

Figure \ref{fig:fig_6} shows the trade network of the suspicious group. The trades with the account theft group in different locations are labeled as ``Group A". Trades with the account theft group at least one time in the same locations are labeled as ``Group B". 
The ``Group A" can be understood as a trade among general users since blacklisted IP addresses are not used, and it does not happen in the locations used by the account theft group. In the case of `(e)' and `(f)' in the ``Group B", it is believed that they have nothing to do with the abused accounts since original users trade for a week through IP addresses as they have. `(g)' is conceived to be used by the account theft group as a new IP address not used for a week is used and the trade happens in the location, `D' where the group trades. In other words, the Figure \ref{fig:fig_4} puts together `(d)' of the Figure \ref{fig:fig_5} and `(g)' of the Figure \ref{fig:fig_6} as trade connections with U21 at its center.   

This study finds that the account theft collects items or game money by using victim accounts and move them to somewhere for final delivery.\\

\section{Conclusion}
\label{section:conclusion}
Account theft is the most serious threat which can damage to both game companies and game users. 
In this study, we propose the method to reveal account thieve' pattern through data mining and action sequence analysis.  
The proposed detection method shows a higher detection rate over `less than 1-minute' type which was classified as the `prelude' acts before the actual account theft proceed. By using this pattern, game companies can pro-actively detect and stop the account theft attempts.  
Also, we conduct trade network analysis to reveal the victims accounts stolen by account thieves are correlated with GFGs. 
In the future, we will adopt more  features to increase the accuracy. We will also consider to apply this model to another MMORPG. We believe our method can apply to many online games which have detailed game logs including transactions and logon events. 

\bibliographystyle{acm}
\bibliography{netgames}

\end{document}